
\input harvmac
\Title{\vbox{\baselineskip12pt\hbox{UGVA-DPT 1994/11-868}\hbox{DFUPG 98/94}}}
{\vbox{\centerline {Self-Duality and Oblique Confinement}
\vskip2pt\centerline{in Planar Gauge Theories}}}
\centerline{M. C. Diamantini and P. Sodano}
\medskip
\centerline{I.N.F.N. and Dipartimento di Fisica, Universit\`a di Perugia}
\centerline{via A. Pascoli, I-06100 Perugia, Italy}
\bigskip
\centerline{C. A. Trugenberger\foot{Supported by a Profil 2 fellowship of the
Swiss National Science Foundation}}
\medskip
\centerline{D\'epartement de Physique Th\'eorique, Universit\'e de
Gen\`eve}
\centerline{24, quai E. Ansermet, 1211 Gen\`eve 4, Switzerland}
\vskip .3in
We investigate the non-perturbative structure of two planar $Z_p \times Z_p$
lattice gauge models and discuss their relevance
to two-dimensional condensed matter
systems and Josephson junction arrays.
Both models involve two compact $U(1)$ gauge fields with Chern-Simons
interactions, which break the symmetry down to $Z_p \times Z_p$.
By identifying the relevant
topological excitations (instantons) and their interactions we determine the
phase structure of the models.
Our results match observed quantum phase transitions in Josephson junction
arrays and suggest also the possibility of {\it oblique confining
ground states} corresponding to quantum Hall regimes for either charges
or vortices.

\Date{December 1994}

\newsec{Introduction}
Topological excitations play a fundamental role in gauge theories
with a compact gauge group \ref\pol{For a review see:
A. M. Polyakov, "Gauge Fields and Strings", Harwood Academic
Publishers, Chur (1987).} . Instantons, i.e. topological
saddle-point configurations in Euclidean space-time can lead
to drastic modifications of the perturbative behaviour of
a theory, like confinement and a mass for the gauge fields.
Typically, the non-perturbative phase structure of a theory is
determined by the condensation (or lack thereof) of such
topological configurations in the ground state.

In the case of Abelian gauge theories, the compact $U(1)$ group
can be automatically obtained by spontaneous breakdown of a
compact, non-Abelian gauge group. In this case, the ultraviolet
cutoff determining the instanton scale is provided by the mass
of the gauge fields corresponding to the broken symmetry generators.
Alternatively, one can formulate the $U(1)$ model on a lattice,
with the gauge fields being phases of link variables \ref\kog{For
a review see: J. B. Kogut, Rev. Mod. Phys. 55 (1983) 775.} .
In this case,
the instanton scale is provided by the lattice spacing.
This formulation is also particularly suited to study models
where the compact $U(1)$ group is broken down to a discrete
gauge group $Z_p$.

In (3+1) dimensions, $Z_p$ lattice gauge theories display a very
interesting phase structure \ref\eli{S. Elitzur, R. Pearson and
J. Shigemitsu, Phys. Rev. D19 (1979) 3698; D. Horn, M. Weinstein
and S. Yankelowicz, Phys. Rev D19 (1979) 3715; A. Guth, A. Ukawa
and P. Windey, Phys. Rev. D21 (1980) 1013.} . There are two
types of string-like topological excitations carrying electric
and magnetic quantum numbers respectively. The models are
{\it self-dual} in the sense that the partition function is
invariant under the duality transformation exchanging the
electric and magnetic excitations and substituting the coupling
constant with its inverse. Self-duality is reflected in the
phase structure of the models: there is a Higgs
phase, when the electric excitations condense in the ground
state and a confinement phase when the
magnetic excitations condense in the ground state. In these phases,
magnetic and electric charges different from multiples of $p$
are confined, respectively; the mechanism leading to confinement
is thus the dual Meissner effect \ref\tma{S. Mandelstam,
in "Extended Systems in Field Theory" , J. L. Gervais and
A. Neveu eds., Phys. Rep. C23 (1976); G. 't Hooft, in
"High Energy Physics: Proceedings of the EPS International
Conference, Palermo (June 1975)", A. Zichichi ed., Editrice
Compositori, Bologna (1976).} . The phase
transition occurs at the self-dual point, where the coupling constant
is invariant under the duality transformation. The photon is massive
in both phases. For large enough $p$, however, the Higgs and
confining phases can be
separated by a Coulomb phase, in which neither topological excitation
condenses in the ground state and the photon is massless.

The complexity of the phase structure is highly increased if
a topological $\theta $-term is added to the action \ref\cra{J. L.
Cardy and E. Rabinovici, Nucl. Phys. B205 [FS5] (1982) 1; J. L.
Cardy, Nucl. Phys. B205 [FS5] (1982) 17; A. Shapere and F. Wilczek,
Nucl. Phys. B320 (1989) 669.} .
In this case, the
magnetic excitations carry also electric charge \ref\wit{E. Witten,
Phys. Lett. 86B (1979) 283.} . As a consequence, in addition to
the previously described Higgs, confinement and Coulomb phases, we
can have new phases characterized by the condensation of topological
excitations carrying both electric and magnetic quantum numbers.
In this phases, only particles carrying electric and magnetic
quantum numbers in the same ratio as in the condensate emerge as
non-confined, physical particles. These phases are therefore called
{\it oblique confinement} phases \ref\hoo{G. 't Hooft, Nucl. Phys.
B190 [FS3] (1981) 455.} .

In (2+1) dimensions, the relevant instanton configurations of
a compact $U(1)$ gauge theory are point-like \ref\pop{A. M.
Polyakov, Phys. Lett. 59B (1975) 82, Nucl. Phys. B120 (1977) 429.}
\ and coincide with the familiar Dirac magnetic monopoles
\ref\gol{For a review see: P. Goddard and D. Olive, Rep. Prog.
Phys. 41 (1978) 91.} \ of three-dimensional Minkowski space.
These instantons lead to confinement of the fundamental charges
of the model and
endow the photon with a non-perturbative mass. It is however
known \ref\pis{R. D. Pisarski, Phys. Rev. D34 (1986) 3851;
I. Affleck, J. Harvey, L. Palla and G. W. Semenoff, Nucl. Phys.
B328 (1989) 575.} \ \ref\dst{M. C. Diamantini, P. Sodano and
C. A. Trugenberger, Phys. Rev. Lett. 71 (1993) 1969.} \ that
the monopoles are linearly confined themselves, if a topological
Chern-Simons term is added to the action. In this case, the relevant
topological configurations are string-like: closed strings or
open strings with a monopole-antimonopole pair at their ends.

In this paper, we investigate two lattice $Z_p \times Z_p$ models in
(2+1) dimensions \foot{Related planar $Z_p$ theories where recently
considered in \ref\bai{F. A. Bais, P. van Driel and M. de Wild Propitius,
Phys. Lett. B280 (1992) 63, Nucl. Phys. B393 (1993) 547; F. A. Bais,
A. Morozov and M. de Wild Propitius, Phys. Rev. Lett. 71 (1993) 2383.}}
which exhibit analogous features to their
(3+1)-dimensional counterparts, namely self-duality and
oblique confinement. Both models involve two compact
$U(1)$ gauge fields coupled via a mixed Chern-Simons term.
Thus, magnetic flux for one gauge field plays the role of the
charge coupled to the other. It is this mixed Chern-Simons
coupling, which breaks both $U(1)$ gauge groups down to
discrete groups.
We study non-perturbative features of these models by
identifying the
relevant topological configurations and their interactions by a
duality transformation \ref\sav{For a review see: R. Savit, Rev. Mod.
Phys. 52 (1980) 453} . Contrary to the case of Maxwell-Chern-Simons
theory, there are no difficulties \ref\esu{D. Eliezer and G. W.
Semenoff, Ann. Phys. (N.Y.) 217 (1992) 66.} \ \dst \ in formulating
a compact lattice gauge model, since the Chern-Simons term is a
mixed one.

There are two types of string-like (Euclidean) topological excitations
corresponding to the two available charge currents
(or magnetic fluxes):
contrary to the (3+1)-dimensional $Z_p$ models, these can be
open, in which case they terminate on monopole-antimonopole
pairs. These monopoles describe tunneling events leading
to the creation (or destruction) of $p$ localized charges
(or magnetic fluxes). The two charges are indeed conserved
only modulo $p$, due to the discrete gauge symmetry
$Z_p\times Z_p$ \ref\kle{K. Lee, Nucl. Phys. B373 (1992) 735.} .
Local gauge invariance is not violated, since the topological
excitations couple to the gauge fields only through their curls,
which represent topologically conserved currents.
The various phases of our models are characterized by the condensation
(or lack thereof) of these topological excitations in the ground
state. The order parameters distinguishing the different phases
are the Wilson loop \pol \ expectation values for the two gauge
fields. Contrary to the (3+1)-dimensional models, the photon is
massive in all possible phases. This photon mass is a topological
one, originating from the mixed Chern-Simons coupling between the
two gauge fields.

In addition to their intrinsic field theoretic interest, the
models we study are of relevance as effective field theories for
two-dimensional condensed matter systems. Indeed, planar gauge
fields play an important role in describing the low-energy degrees
of freedom for such systems. The key point is that in (2+1)
dimensions, a conserved matter current $j^{\mu }$ can always be
represented in terms of a pseudovector Abelian gauge field $B_{\mu }$
as \eqn\rcc{j^{\mu } \propto \epsilon ^{\mu \alpha \nu}
\partial _{\alpha } B_{\nu } \ .}
This current is usually taken to represent the low-energy
matter fluctuations above a given ground-state. The effective
theory governing the dynamics of these fluctuations can then be
written in terms of $B_{\mu }$ as a {\it gauge theory}. The behaviour
of the matter fluctuations is dominated by the lowest
dimension term appearing in the gauge field action. Naturally,
relativistic invariance does not play any crucial role
in these applications.
However, relativistic gauge theories provide a framework for
studying the relevant physical phenomena, just as the Abelian
Higgs model describes the essential features of Landau-Ginzburg
effective theories of superconductivity \ref\raj{For a review
see: R. Rajaraman, "Solitons and Instantons", North-Holland,
Amsterdam (1987).} .
In these applications to planar condensed matter systems, the
gauge symmetry is {\it compact}, reflecting the underlying
lattice structure of the original microscopic model.

The second gauge field in our models lends itself to two possible
interpretations.
In applications to Josephson junction arrays
\ref\faz{U. Eckern and A. Schmid, Phys. Rev. B39 (1989) 6441;
R. Fazio and G. Sch\"on, Phys. Rev. B43 (1991) 5307;
R. Fazio, U. Geigenm\"uller and G. Sch\"on, contribution in
"Quantum Fluctuations in Mesoscopic and Macroscopic Systems",
H. A. Ceredeira ed., World Scientific, Singapore (1991).} \ we take
it to encode the vortex dynamics according to an
equation analogous to \rcc . Therefore, also the second gauge group
is compact. In applications to generic planar condensed matter systems,
the second gauge field is taken to describe electromagnetic
fluctuations coupled to the low-energy matter excitations. The resulting
models are {\it effective gauge theories}, valid on scales much larger
than $1/\Lambda $, with $\Lambda $ the ultraviolet cutoff above which
higher-lying matter excitations become important. In order to incorporate
the dynamics of magnetic vortices (on scales $1/\Lambda $)
in these effective theories, also the
electromagnetic gauge field has to be taken as a {\it compact} variable.
The same happens in the Abelian Higgs model, if we neglect completely
the radial fluctuations of the Higgs field \ref\frash{E. Fradkin and
S. H. Shenker, Phys. Rev. D19 (1979) 3682.} . Thus, we shall always
consider both gauge fields as compact variables.

This paper is organized as follows. In section 2, we formulate
the continuum version of the two models in Minkowski space-time
and discuss
their relevance to planar condensed matter systems and Josephson junction
arrays. In section 3 we introduce our lattice notation, with
particular emphasis on the lattice version of the Chern-Simons
operator. Sections 4 and 5 are devoted to the lattice formulation
of the two models, with compact gauge symmetries,
and to the analysis of their phase structure.
We shall draw our conclusions in section 6.

\newsec{Formulation of the models}

\subsec{Self-dual gauge model in (2+1) dimensions}
The first model we consider involves a vector gauge field
$A_{\mu }$ and a pseudovector gauge field $B_{\mu }$ and is
defined by the Minkowski space-time Lagrangian (units $c=1$ and $\hbar =1$)
\eqn\dla{\eqalign{{\cal L}_{SD} &= {-1\over 4e^2} F_{\mu \nu}
F^{\mu \nu} + {\kappa \over 2\pi } A_{\mu }\epsilon^{\mu \alpha
\nu }\partial _{\alpha }B_{\nu } + {-1\over 4g^2} f_{\mu \nu}
f^{\mu \nu } \cr
&= {-1\over 2e^2} F_{\mu }F^{\mu } + {\kappa \over 2\pi } A_{\mu }
\epsilon ^{\mu \alpha \nu }\partial _{\alpha }B_{\nu } +
{-1\over 2g^2} f_{\mu }f^{\mu } \ ,\cr }}
where the field strengths and their duals are
given by
\eqn\fsd{\eqalign{F_{\mu \nu} &\equiv \partial _{\mu }A_{\nu }
-\partial _{\nu }A_{\mu }\ ,\qquad \qquad F^{\mu } \equiv
{1\over 2}\epsilon ^{\mu \alpha \beta} F_{\alpha \beta}\ ,\cr
f_{\mu \nu} &\equiv \partial _{\mu } B_{\nu } - \partial _{\nu }
B_{\mu }\ ,\qquad \qquad f^{\mu } \equiv {1\over 2}
\epsilon ^{\mu \alpha \beta} f_{\alpha \beta} \ .\cr }}
Note that the mixed Chern-Simons coupling does not violate
the discrete symmetry of parity, due to the pseudovector
character of the gauge field $B_{\mu }$.

The coupling constants $e^2$ and $g^2$ have dimension mass,
whereas the coefficient $\kappa $ of the mixed Chern-Simons term
is dimensionless. In the continuum theory with non-compact gauge
fields one could set $\kappa =1$
and $e=g$ by a rescaling of both gauge fields. This is no more
possible if the gauge fields are compact variables with a fixed
periodicity. This is why we prefer to keep all coupling constants
explicit.

The action of the model \dla \ is separetely invariant under
the two Abelian gauge transformations
\eqn\gtr{\eqalign{A_{\mu }\ &\to \ A_{\mu }+\partial _{\mu }
\lambda \ ,\cr
B_{\mu } \ &\to \ B_{\mu }+\partial _{\mu } \omega \ .\cr }}
The corresponding currents $f^{\mu }$ and $F^{\mu }$ are
topologically conserved. Thus, magnetic flux for one gauge field
plays the role of the conserved charge coupled to the other.
Moreover, the action corresponding to \dla \ is also invariant
under the {\it duality transformation}
\eqn\dtr{\eqalign{A_{\mu } \ &\longleftrightarrow \ B_{\mu } \ ,\cr
e \ &\longleftrightarrow \ g \ .\cr }}

The model \dla \ has originally been proposed \ref\mav{A. Kovner
and B. Rosenstein, Phys. Rev. B42 (1990) 4748; G. W. Semenoff
and N. Weiss, Phys. Lett. B250 (1990) 117; N. Dorey and N. E.
Mavromatos, Phys. Lett. B250 (1990) 107, Nucl. Phys. B386
(1992) 614.} \
as an effective theory of planar superconductivity without
parity violation. In this application, the conserved current
\eqn\dcu{j^{\mu } \equiv {\kappa\over 2\pi } \ \epsilon^{\mu \alpha
\nu }\partial _{\alpha }B_{\nu } \ ,}
describes {\it matter} fluctuations about a given superconducting
ground-state. These matter fluctuations can be thought of \mav \
as fermion bound states of an underlying microscopic model. In this
case, the coupling constant $g^2$ sets the scale for the pairing gap
$\Delta $: $g^2=12 \pi \Delta /\kappa ^2 $.
The last term in \dla \ represents the kinetic
term for the matter: in (2+1) dimensions this
can be written in terms of the effective gauge field $B_{\mu }$
and describes a single, massless scalar field. This is minimally
coupled to planar photons, whose kinetic term is given by
the first term in \dla .

The particle content of \dla \ can be easily exposed by the
linear transformation
\eqn\ltr{\eqalign{A_{\mu } &= \sqrt{e\over g} \left(
a_{\mu }+b_{\mu } \right) \ ,\cr
B_{\mu } &= \sqrt{g\over e} \left( a_{\mu } - b_{\mu } \right) \ .\cr }}
In terms of the new variables $a_{\mu }$ and $b_{\mu }$, the Lagrangian
\dla \ describes a free theory,
\eqn\frt{{\cal L}_{SD} = {-1\over 2eg} G_{\mu \nu }G^{\mu \nu } +
{\kappa \over 2\pi } a_{\mu }\epsilon ^{\mu \alpha \nu } \partial
_{\alpha } a_{\nu } + {-1\over 2eg} g_{\mu \nu }g^{\mu \nu } -
{\kappa \over 2\pi } b_{\mu }\epsilon ^{\mu \alpha \nu }
\partial _{\alpha } b_{\nu }\ ,}
where $G^{\mu \nu }$ and $g^{\mu \nu }$ are the field strengths for the
gauge fields $a_{\mu }$ and $b_{\mu }$, respectively.
This transformation exposes the mechanism of superconductivity: the
original spin 0 and massless photon "absorbs" the matter degree of
freedom, thereby turning into a parity and spin ($\pm 1$) doublet
with a topological mass \ref\djt{R. Jackiw and S. Templeton, Phys. Rev.
D23 (1981) 2291; J. Schonfeld, Nucl. Phys. B185 (1981) 157; S. Deser,
R. Jackiw and S. Templeton, Phys. Rev. Lett. 48 (1982) 975, Ann. Phys.
(N.Y.) 140 (1982) 372.}
\eqn\mas{m = {|\kappa |eg \over 2 \pi } \ .}

As was pointed out in \mav , the photon kinetic term has to be
modified for potential applications of the model to real
quasi-planar high-$T_C$ materials. In these applications,
the dynamics of matter is taken as (2+1)-dimensional, while
the electromagnetic field is the real (3+1)-dimensional one.
The proper way to describe the coupling of (3+1)-dimensional
electromagnetic fields to charges and currents confined to a plane
was derived in \ref\dag{E. Dagotto, A. Kocic and J. Kogut, Phys. Rev.
Lett. 62 (1989) 1083.} :
\eqn\mpr{{\cal L}_{SDP} = {-1\over 4e^2} F_{\mu \nu } {1\over
\sqrt{\partial ^2}} F^{\mu \nu } + {\kappa \over 2\pi } A_{\mu }
\epsilon ^{\mu \alpha \nu }\partial _{\alpha }B_{\nu } +
{-1\over 4g^2} f_{\mu \nu }f^{\mu \nu } \ .}
Here, $\partial ^2 \equiv \partial _{\mu }\partial ^{\mu }$ and
$F^{\mu \nu }$ represents the component of the magnetic field
perpendicular to the plane and the in-plane components of the electric
field. With this modification, $e^2$ is the usual, dimensionless
coupling constant of (3+1)-dimensional electromagnetism. It is easy
to convince oneself that \mpr \ leads to a $1/r$ potential between
static charges. The effective photon mass of this modified model is easily
obtained by integrating out the gauge field $B_{\mu }$ and computing
the resulting photon propagator:
\eqn\mma{\mu ={{\kappa ^2 e^2 g^2} \over 4\pi ^2 \ .}}

In addition to its relevance as an effective theory of planar
superconductivity, the model \dla \ can also be interpreted as
a gauge theory formulation of a planar two-fluid model of coupled
charges and vortices. In this application, the gauge field $A_{\mu }$
provides an effective description of the vortices via the identification
of
\eqn\voc{\Phi ^{\mu } ={\kappa \over 2\pi }\ \epsilon^{\mu \alpha
\nu }\partial _{\alpha } A_{\nu } }
with the (pseudovector) vortex current. The mixed Chern-Simons term
describes then both the Lorentz force exherted by the vortices on the
charges and the Magnus force \ref\mag{For a review see: V. L. Streeter,
"Fluid Dynamics", McGraw-Hill, New York (1948); P. D. McCormack and
L. Crane, "Physical Fluid Dynamics", Academic Press, New York (1973).} \
exherted by the charges on the vortices. Indeed, the Magnus force is
completely analogous to the Lorentz force: vorticity plays the role of
electric charge and fluid density plays the role of the magnetic field.

Charge-charge and vortex-vortex interactions are best exposed in
the Coulomb gauge Hamiltonian derived from \dla .
This can be written entirely
in terms of the charge and vortex currents $j^{\mu }$ and $\Phi ^{\mu }$ :
\eqn\haf{\eqalign{H &= \int d^2 {\bf x}
\ \left\{ j^0 \left( {e^2\over 2} {1\over
-\nabla ^2} + {2\pi ^2 \over \kappa ^2 g^2} \right) j^0 + \Phi ^0
\left( {g^2\over 2} {1\over -\nabla ^2} +{2\pi ^2\over \kappa ^2 e^2}
\right) \Phi ^0 \right\} \cr
&+ \int d^2 {\bf x} \ \left\{ {2\pi ^2\over \kappa ^2 g^2} \ {\bf j}_L^2
+{2\pi ^2 \over \kappa ^2 e^2} \ {\bf \Phi }^2_L \right\} \ ,\cr }}
where $j^i_L$ and $\Phi ^i_L$ denote the longitudinal components
of the charge and vortex current densities, respectively.
As expected, both the charges and the vortices are subject to
long-range Coulomb interactions; there are no charge-vortex
interactions other than the Lorentz and Magnus forces mentioned above
and these do not contribute to the Hamiltonian. The last two terms
in $H$ represent the kinetic terms for charge and vortex motion,
respectively.

The coupled Coulomb gas of charges and vortices described by
\dla \ and \haf \ is reminiscent of well known statistical
mechanics systems, namely Josephson junction arrays \faz .
In order to make further contact with these systems, let us
formulate the Hamiltonian \haf \ on a square lattice with lattice
spacing $l$. To this end we consider the charges and vortices as
variables defined on the sites of the lattice (denoted by ${\bf x}$),
whereas the currents are associated with the links (denoted by
$({\bf x}, i)$). Introducing a lattice is actually not sufficient
to completely regularize the problem. Indeed, in two spatial dimensions,
the lattice Green function
$G({\bf x}-{\bf y})$ representing
the inverse lattice Laplace operator \ref\kad{L. P. Kadanoff, J. Phys. A:
Math. Gen. 11 (1978) 1399.} \ is still
logarithmically divergent for ${\bf x}-{\bf y}=0$
and a further regularization is needed.
This leaves the ambiguity of a finite subtraction constant $\alpha $.
We thus obtain the following lattice Hamiltonian:
\eqn\lah{\eqalign{ H_L &= \sum _{\bf x, \bf y} q_{\bf x} \ {e^2\over 2}
V({\bf x}-{\bf y}) \ q_{\bf y}
+ \sum_{\bf x, \bf y} \phi _{\bf x} \ {g^2\over 2} V({\bf x}-{\bf y})
\ \phi _{\bf y} \cr
&+ \sum _{\bf x, i} {e^2\over 2(ml)^2} \ q_{\bf x}^2
{v_q^i}^2 + \sum _{\bf x , i} {g^2\over 2(ml)^2}
\ \phi _{\bf x}^2 {v_{\phi}^i}^2 \ .\cr }}
Here, $m$ is the topological mass \mas ,
$q_{\bf x}$ and $\phi _{\bf x}$ denote the charge and vortex
numbers at site ${\bf x}$ respectively whereas $v^i_q$ and $v^i_{\phi }$
label the charge and vortex velocities on the link $({\bf x}, i)$.
The Green function $V({\bf x}-{\bf y})$ is a lattice kernel with the
property $V(0)=\left[ (1/(ml)^2) -\alpha \right] $ and behaving
as $-({\rm log} |{\bf x}-{\bf y}| /2\pi)-\alpha $ at distances large compared
to
the lattice spacing.

The first two terms in $H_L$ describe two Coulomb gases of charges and
vortices, respectively. The other two terms in $H_L$ represent kinetic
terms for these charges and vortices. They imply the
following masses for charges $q$ and vortices $\phi $:
\eqn\mcv{\eqalign{m_q &= {q^2 e^2 \over (ml)^2} \ ,\cr
m_{\phi } &= {\phi ^2 g^2\over (ml)^2} \ . \cr }}

With an appropriate choice of the subtraction constant $\alpha $
and for $\kappa =2$ (representing the charge of Cooper pairs)
the Hamiltonian \lah \ reduces essentially to the Hamiltonian of a
Josephson junction array
\faz \ upon identifying the charging energy $E_C$ and the
Josephson coupling $E_J$ as
\eqn\ide{E_C={e^2\over 4}\ ,\qquad \qquad E_J={g^2 \over 2 \pi ^2} \ .}
With this identification,
the topological mass \mas \ (for $\kappa =2$) coincides with
the plasma frequency $\sqrt{8E_CE_J}$ of the array.

The difference between $H_L$ and the Hamiltonian describing the arrays
lies in the kinetic term for the vortices (last
term in $H_L$), which is absent in the latter.
It is the absence of this term which breaks the
perfect duality \foot{In terms of the array variables $E_C$ and $E_J$, the
self-dual point $g/e=1$ is given by $E_J/E_C = 2/\pi ^2$ .}
of \lah \ in the real systems \faz .
The vortex kinetic term in our model is connected to the presence
of a doublet of propagating degrees of freedom (see \frt ). Indeed,
we could get rid of it by simply projecting out the transverse components
of the electric field for $A_{\mu }$ from the action and the Hamiltonian.
This would leave us with a single propagating degree of freedom of mass
\mas , representing essentially the plasmons in the array.
We don't expect the additional vortex kinetic term to induce drastic
modifications in the regimes $e/g \ll 1$ ($E_C/E_J \ll 1$) and
$e/g \gg 1$ ($E_C/E_J \gg 1$), where either charges or vortices clearly
dominate the dynamics. However, it is harder to estimate the influence
of the additional term in the intermediate region $e/g \simeq 1$.

\subsec{Oblique confining model in (2+1) dimensions}
In (3+1) dimensions, the addition of a topological $\theta $-term
to the action of lattice $Z_N$ gauge models leads to the appearance
of new, oblique confinement phases, characterized by the condensation
of topological excitations carrying
both electric and magnetic charges \cra .
It is therefore natural to investigate if the same phenomenon can take
place in (2+1) dimensions. In (2+1) dimensions, the natural topological
term to add to the Lagrangian \dla \ is a Chern-Simons term involving
only one of the gauge fields, say $B_{\mu }$. However, the model so
obtained does not lead to a simple dual (generalized) Coulomb gas
representation \sav \ on a cubic lattice,
due to the usual difficulties \esu \ \dst \ in inverting
the lattice Chern-Simons
operator. As we show below, nonetheless,
one can get rid of this problem if a
further coupling $F^{\mu }f_{\mu }$ is added to the Lagrangian.
We consider thus a model defined by the Lagrangian density
\eqn\ocm{{\cal L}_{OC} = {-1\over 2e^2} F_{\mu }F^{\mu } +
{\kappa \over 2\pi }A_{\mu }\epsilon ^{\mu \alpha \nu }\partial _{\alpha }
B_{\nu } - \lambda F_{\mu }f^{\mu } +{-1\over 2g^2} f_{\mu }f^{\mu }
+{\eta \over 2\pi } B_{\mu }\epsilon ^{\mu \alpha \nu }\partial _{\alpha }
B_{\nu } \ .}
There are two new coupling constants: $\eta $ is dimensionless, whereas
$\lambda $ has dimension ${\rm mass}^{-1}$. Both the two new
terms violate parity; gauge invariance is clearly maintained, whereas
na\"ive self-duality is broken by the Chern-Simons term.

If we maintain the interpretation \dcu \ of $(\kappa / 2\pi )
\epsilon ^{\mu \alpha \nu }\partial _{\alpha }B_{\nu }$ as a conserved
current describing matter fluctuations about the ground state of an
underlying statistical mechanics model, the additional Chern-Simons
term describes a non-local Hopf interaction for this current. The matter
degree of freedom is then a topologically massive field \djt \ of
mass $\eta g^2/\pi $ and spin $s=\eta /|\eta |$.
The additional coupling $f_{\mu } F^{\mu }$ has the form of a
(relativistic) Pauli interaction; correspondingly, the new coupling
$\lambda $ can be viewed as an intrinsic magnetic moment for the matter.
In our model, we shall fix this new parameter as follows.
By integrating out the electromagnetic gauge field $A_{\mu }$,
we obtain an effective theory for the matter degree of freedom:
\eqn\eff{{\cal L}^B_{\rm eff} =-{1\over 2} \left( {1\over g^2} -e^2
\lambda ^2 \right) f_{\mu }f^{\mu } + {e^2\kappa ^2\over 8\pi ^2}
B_{\mu }\left( \delta ^{\mu \nu } -{ \partial ^{\mu }\partial ^{\nu }
\over \partial ^2 }\right) B_{\nu } +{{\eta - \kappa \lambda e^2}\over
2\pi } B_{\mu }\epsilon ^{\mu \alpha \nu }\partial _{\alpha }B_{\nu } \ .}
For generic $\lambda $, this theory contains both a Higgs mass and a
topological Chern-Simons mass. We shall fix $\lambda $ by the requirement
that the induced Chern-Simons term cancels exactly the bare one, so that
only the Higgs mass survives:
\eqn\lac{\lambda = {\eta \over \kappa e^2} \ .}
For this choice of $\lambda $, the interaction with electromagnetic
fluctuations is able to lift the frustration in the matter dynamics
represented by the current-current Hopf interaction (Chern-Simons term).
As we shall show in section 5, it is also this cancellation of
the bare and induced Chern-Simons terms, which allows a simple lattice
Coulomb gas representation for the topological excitations of the model.
With the value \lac \ for $\lambda $, \eff \ describes a parity
doublet of excitations with spin $\pm 1$ \ref\bin{B. Binegar, J. Math.
Phys. 23 (1982) 1511.} \ and mass
\eqn\mms{M= {{eg\kappa \over 2\pi} \over \sqrt{1-{\eta ^2 g^2 \over
\kappa ^2 e^2}}} \ .}
In order to avoid tachyonic excitations, the remaining parameters
of the theory must satisy the condition $\eta g/\kappa e\le 1$.

When \lac \ is satisfied, our model \ocm \ is related to the self-dual
model introduced in the previous section by a simple transformation
of parameters. This is immediately clear, once it is realized that the
Lagrangian \ocm \ can be rewritten as
\eqn\nfo{\eqalign{{\cal L}_{OC} &= - {1\over 2e^2}
\left( F_{\mu } +{\eta \over \kappa}
f_{\mu } \right) \left( F^{\mu }+{\eta \over \kappa}f^{\mu } \right)
+{\kappa \over 2\pi } \left( A_{\mu }+{\eta \over \kappa }B_{\mu } \right)
\epsilon ^{\mu \alpha \nu }\partial _{\alpha }B_{\nu } \cr
&-{1\over 2g^2} \left( 1-{{\eta ^2 g^2} \over {\kappa ^2 e^2}} \right)
f_{\mu }f^{\mu } \ .\cr }}
By introducing a new gauge field $C_{\mu }$ and a new coupling constant
$g'$ defined as
\eqn\ngc{C_{\mu } \equiv A_{\mu } +{\eta \over \kappa } B_{\mu } \ ,
\qquad \qquad g' \equiv {g\over \sqrt{1-{{\eta ^2 g^2} \over {\kappa ^2 e^2}}}}
\ , }
we recover exactly the self-dual model \dla . Correspondingly, we can
express the mass $M$ as
\eqn\mapr{M= m(e, g') \ .}
We conclude therefore that
our model \ocm \ has a {\it hidden duality symmetry} when \lac \ is
satisfied.
Note that the new self-dual point $e/g= \sqrt{1+\eta^2 /\kappa ^2}$ lies
in allowed range of parameters $g/e <\kappa /\eta $ for all values of
$\kappa $ and $\eta $.

The model \ocm \ is also related to known planar condensed matter
systems. Indeed, the theory with Lagrangian
\eqn\cif{{\cal L}_{CIF} = {\kappa \over 2\pi } A_{\mu }
\epsilon ^{\mu \alpha \nu } \partial _{\alpha }B_{\nu }
+{\eta \over 2\pi }B_{\mu } \epsilon ^{\mu \alpha \nu }\partial _{\alpha }
B_{\nu } \ ,}
has been proposed \ref\fwz{J. Fr\"ohlich and A. Zee, Nucl. Phys.
B364 (1991) 517; X.-G. Wen and A. Zee, Phys. Rev. B46 (1993) 2290.}\
as the effective field theory describing the long distance behaviour
of {\it chiral incompressible fluids}
\ref\zee{For a review see: A. Zee, "Long Distance Physics of
Topological Fluids", Progr. Theor. Phys. Supp. 107 (1992) 77.} .

The presence of the Chern-Simons term as the dominant kinetic term
for matter fluctuations reflects an either explicit or spontaneous
breakdown of the discrete $P$ and $T$ symmetries in the underlying
microscopic model.

For $2\eta =$ even integer, the effective field theory \cif \ describes
the long-distance physics of {\it chiral spin liquids} \ref\wwz{X. G.
Wen, F. Wilczek and A. Zee, Phys. Rev. B39 (1989) 11413.} ; in this
case the $P$ and $T$ symmetries are spontaneously broken. For
$2\eta =$ odd integer, the same theory describes the long distance
physics of {\it Laughlin's incompressible quantum fluids}, which
are the matter ground states at the plateaus of the quantum Hall
effect \ref\gip{For a review see: "The Quantum Hall Effect", R. E.
Prange and S. M. Girvin eds., Springer Verlag, New York (1990).} .
In this case, the $P$ and $T$ symmetries are explicitly broken
by the external magnetic field and $\nu = 1/2\eta $ plays the
role of the filling fraction \ref\sos{G. W. Semenoff and
P. Sodano, Phys. Rev. Lett.  57 (1986) 1195.} . This can be easily
seen by integrating out the matter degree of freedom $B_{\mu }$,
to obtain an effective action for the gauge potential $A_{\mu }$:
\eqn\aef{S^A_{\rm eff} = \int d^3 x\ {-\kappa ^2 \over 8\pi \eta } \
A_{\mu }\epsilon ^{\mu \alpha \nu } \partial _{\alpha } A_{\nu } \ .}
The induced current is then given by the usual expression
\eqn\icu{j^{\mu }_{\rm in} \equiv {\delta \over \delta A_{\mu }}
S^A_{\rm eff} = -{\kappa ^2 \over 4\pi \eta } \ \epsilon ^{\mu \alpha
\nu }\partial _{\alpha }A_{\nu } \ .}
The matter current induced by an electric field $E^i$ is thus given by
\eqn\hac{j^i_{\rm in} = {\kappa ^2\over 2\pi } \ {1\over 2\eta }\ \epsilon
^{ij}
E^j \ ,}
which we recognize as the {\it Hall current} for an incompressible liquid
of particles of charge $\kappa $ (in units of the fundamental charge)
and of filling fraction $1/2\eta $.

The effective field theory \cif \ describes the incompressible quantum
fluids in the limit of an {\it infinite gap} for matter fluctuations.
Indeed, there are no propagating modes, due to the topological nature
of both terms in the Lagrangian. Our model \ocm \ can be viewed as an
extension of \cif , in which the three possible terms of dimension
$({\rm mass})^4$ coupling the dual field strengths $F^{\mu }$ and
$f^{\mu }$ have been added to the Lagrangian. These can be
interpreted as the next-to-leading terms appearing in a derivative
expansion of a local, relativistic, gauge invariant effective action
for purely planar (also the electromagnetic fluctuations are taken
to be (2+1)-dimensional) incompressible fluids.
They provide {\it dynamics} for the gauge
fields $A_{\mu }$ and $B_{\mu }$, which become propagating degrees of
freedom. The resulting topologically
massive matter mode represents the so called
{\it magnetophonon} \ref\gir{For a review see:
S. M. Girvin, "Collective Excitations", in \gip .} .

With the alternative interpretation of $(\kappa /2\pi )F^{\mu }$ as a
vortex current, instead, we expect \nfo \ to capture the essential
physics of Josephson junction arrays in the presence of
$\eta $ external offset charges per plaquette.
Correspondingly, had we added a Chern-Simons term for $A_{\mu }$,
instead of $B_{\mu }$, we would describe the same systems in presence of
an external magnetic field with $\eta $ fluxes per plaquette.
This conjecture, motivated by the analogy with
the quantum Hall effect, leads to predictions on the $T=0$ phase structure
(see section 5) that might be accessible experimentally.

As in the case of the self-dual model of the previous section, \ocm \ has
to be slightly modified for potential applications to real quantum Hall
samples.
Specifically, one must incorporate again (3+1)-dimensional effects and, in
particular, a $1/r$ interaction between charges. Given the representation
\nfo , this can be achieved by modifying the model to
\eqn\rqh{{\cal L}_{OCP} = -{1\over 2e^2}
\left( F_{\mu } +{\eta \over \kappa }
f_{\mu } \right) {1\over \sqrt{\partial ^2}} \left(F^{\mu } +{\eta \over
\kappa }f^{\mu } \right) +{\kappa \over 2\pi } \left( A_{\mu }+{\eta \over
\kappa } B_{\mu } \right) \epsilon ^{\mu \alpha \nu }\partial _{\alpha }
B_{\nu } -{1\over 2{g'}^2} f_{\mu } f^{\mu } \ ,}
with $e^2$ dimensionless. Note that this modification also changes the
logarithmic potential between matter vortices (described by the vortex
density $\epsilon ^{ij} \partial _i f_j$) to a linear potential.
Moreover, also the dynamics of free matter excitations (magnetophonons)
is slightly
modified. Nonetheless, the limit ${g'}^2\to \infty $ still describes the
limit of an infinite mass.

As emphasized in the introduction, our model does not
reproduce the exact dynamics of fluctuations about
incompressible quantum fluids; however, it incorporates several
essential features of this dynamics
which are absent in \cif , in particular the
existence of a {\it finite gap}
\eqn\gfe{\Delta = \mu (e, g') }
for the excitations.

\newsec{Lattice Chern-Simons term}
As mentioned in the introduction, we would like to investigate the
non-perturbative structure of the models described in the previous
section, when the gauge fields are compact variables. To this end
we shall study the Euclidean partition function of the models on a
cubic lattice with lattice spacing $l$.
Lattice sites are denoted by the vector ${\bf x}$, and the links
between ${\bf x}$ and ${\bf x} + \hat \mu$, $\mu = 1,\dots ,3$, with
$({\bf x}, \mu)$.
The gauge fields $A_{\mu}$ and $B_\mu$ are associated with each link
$({\bf x}, \mu)$, and for a compact gauge theory they have to
be considered as angular variables defined on the interval
$[- \pi/l, \pi/l]$:
\eqn\per{\eqalign{& A_{\mu}({\bf x}) \equiv A_{\mu}({\bf x}) + {2 \pi
n_\mu ({\bf x}) \over l}\ , \qquad n_\mu({\bf x}) \in Z\ , \cr
& B_{\mu}({\bf x}) \equiv B_{\mu}({\bf x}) + {2 \pi k_\mu
({\bf x}) \over l}\ ,\qquad k_\mu({\bf x}) \in Z\ . \cr }}
On the lattice, we define the following forward and backward
derivatives and shift operators:
\eqn\der{\eqalign{d_\mu f({\bf x}) &\equiv {f({\bf x} + \hat \mu l) -
f({\bf x}) \over l}\ , \qquad S_\mu f({\bf x)} \equiv f({\bf x} +
\hat \mu l) \ , \cr
\hat d_\mu f({\bf x}) &\equiv {f({\bf x}) - f({\bf x} - \hat \mu l)
\over l} \ , \qquad \hat S_\mu f({\bf x}) \equiv f({\bf x} - \hat \mu
l) \ , \cr}}
Summation by parts interchanges both the two derivatives and the
two shift operators:
\eqn\din{\eqalign{\sum_{\bf x} f({\bf x}) \ d_\mu g({\bf x}) &= - \sum_{\bf x}
\hat d_\mu f({\bf x}) \ g({\bf x}) \ ,\cr
\sum _{\bf x} f({\bf x}) \ S_{\mu } g({\bf x}) &= \sum_{\bf x} \hat S_{\mu }
f({\bf x}) \ g({\bf x}) \ ,\cr }}
where we have omitted possible surface terms. Gauge transformations are
defined by using the forward lattice derivative,
\eqn\gau{A_{\mu }({\bf x}) \ \to \ A_{\mu }({\bf x}) + d_{\mu }
\lambda ({\bf x}) \ .}

In order to formulate our models on the lattice, we have to face
the problem of defining a lattice version of the Chern-Simons term.
This problem has recently received much attention
\ref\fro{J. Fr\"ohlich and P.-A. Marchetti, Comm. Math. 121
(1989) 177.} \ \ref\lue{E. Fradkin, Phys. Rev. Lett. 63 (1989) 322;
P. L\"uscher, Nucl. Phys. B326 (1989) 557; V. F. M\"uller, Z. Phys.
C47 (1990) 301.} \ \esu \ and consists basically in defining a suitable
analogue of the Chern-Simons operator $\epsilon_{\mu \alpha \nu}
\partial^{\alpha }$.
It is easy to verify that this operator is the square root of the
familiar Maxwell operator: $\epsilon_{\mu \gamma \alpha }\partial ^{\gamma }
\ \epsilon ^{\alpha \delta \nu }\partial _{\delta }= - {\delta _{\mu }}^{\nu }
\partial ^2 +\partial _{\mu }\partial ^{\nu } $.
While in Minkowsky space-time (with discrete space and continuous time)
this problem has been solved by Eliezer and
Semenoff \esu , for the Euclidean version, on a cubic lattice, it turns
out that there is no gauge invariant, local operator whose square
reproduces the lattice Maxwell operator.
In this case we can, however, define
the following two lattice operators \fro :
\eqn\kap{K_{\mu \nu} \equiv S_\mu \epsilon_{\mu \alpha \nu} d_\alpha
\ , \qquad \hat K_{\mu \nu} \equiv \epsilon_{\mu \alpha \nu} \hat
d_\alpha \hat S_\nu \ ,}
where no summation is implied over equal indices $\mu $ and $\nu $.
These operators are both {\it local} and {\it gauge invariant}, in the
sense that they lead to gauge invariant terms when contracted on both
sides \foot{Note that, on the lattice, gauge invariance requires that
kernels are annihilated by $d_{\mu }$ on the right and by $\hat d_{\mu }$
on the left.} with gauge fields:
\eqn\gau{K_{\mu \nu} d_\nu = \hat d_\mu K_{\mu \nu} = 0 \ , \qquad
\hat K_{\mu \nu} d_\nu = \hat d_\mu \hat K_{\mu \nu} = 0 \ ,}
The squares of $K_{\mu \nu }$ and $\hat K_{\mu \nu }$ do not have
any particular meaning; however, the product of the two operators
reproduces the lattice Maxwell operator,
\eqn\squ{K_{\mu \alpha} \hat K_{\alpha \nu} = \hat K_{\mu \alpha}
K_{\alpha \nu} = -\delta_{\mu \nu} \nabla^2 + d_\mu \hat d_\nu \ ,}
where $\nabla^2 \equiv \hat d_\mu d_\mu$ is the three-dimensional,
Euclidean Laplace operator on the lattice.
In analogy to the forward and backward derivatives and shift operators,
also $K_{\mu \nu}$ and
$\hat K_{\mu \nu}$ are interchanged upon summation by parts,
\eqn\ink{\sum _{\bf x, \mu } A_{\mu } K_{\mu \nu } B_{\nu }
=\sum _{\bf x, \mu } B_{\mu }\hat K_{\mu \nu } A_{\nu } \ .}
Both $K_{\mu \nu }$ and $\hat K_{\mu \nu }$ can be used to define
a Chern-Simons term in the lattice action. Hereafter, we choose to use
$K_{\mu \nu}$.

Using $K_{\mu \nu }$ we can also define the lattice dual field strengths as
\eqn\ldf{\eqalign{F_{\mu } &\equiv K_{\mu \nu } A_{\nu } \ ,\cr
f_{\mu } &\equiv K_{\mu \nu } B_{\nu } \ .\cr }}
These are also compact variables, defined on the interval
$[-\pi /l^2, \pi /l^2]$. By using \ink \ and \squ , we easily obtain
the following identity:
\eqn\lmt{\sum_{\bf x, \mu } F_{\mu }^2 =
\sum_{\bf x, \mu } A_{\mu } \left( -\delta _{\mu \nu }\nabla ^2 +
d_{\mu }\hat d_{\nu } \right) A_{\nu } \ ,}
which shows that we can write the lattice Maxwell action simply as
$(l^3/2e^2) \sum_{\bf x, \mu } F_{\mu }^2$.

\newsec{Self-dual model: non perturbative analysis}
\subsec{Lattice formulation and topological excitations}
In order to take into account the periodicity of the gauge fields
$A_\mu$ and $B_\mu$, we introduce four sets of integer link variables
$\{n_{\mu }\}$, $\{l_{\mu }\}$, $\{k_{\mu }\}$ and $\{m_{\mu }\}$, and we
posit the following Euclidean lattice partition function of the
Villain type \sav :
\eqn\lam{\eqalign{Z &= \sum _{{\{n_{\mu }\}, \{l_{\mu }\} }\atop {
\{k_{\mu }\} , \{ m_{\mu }\} }} \int _{-\pi \over l}^{\pi \over l}
 {\cal D} A_{\mu } {\cal D} B_{\mu } \ {\rm exp}(-S) \ ,\cr
S &= \sum_{{\bf x}, \mu }\  {l^3\over 2e^2} \left( F_{\mu } +{2\pi
\over l^2} n_{\mu } \right) ^2 - i{l^3 \kappa \over 2\pi } \left(
A_{\mu } + {2\pi \over l} l_{\mu } \right) K_{\mu \nu } \left( B_{\nu
} +{2\pi \over l} m_{\mu } \right) \cr
&\ \ \ \ \ \ \ \ \ +{l^3\over 2g^2}
\left( f_{\mu } + {2\pi \over l^2} k_{\mu } \right)
^2 \ ,\cr }}
where we have introduced the notation ${\cal D} A_\mu \equiv
\prod_{{\bf x}, \mu} dA_\mu({\bf x})$.
This partition function is clearly invariant under the shifts \per ,
since these can be reabsorbed by a redefinition of the integer link
variables. For $\kappa =0$, \lam \ reduces to the sum of two uncoupled
copies of the Villain action for compact $U(1)$ gauge fields studied by
Polyakov \pol . In this case, the relevant topological excitations are
point-like monopoles, one type for each gauge field. The mixed Chern-Simons
coupling between the two gauge fields requires the introduction of two
additional integer link variables, in order to maintain the periodicity
of the full action.

These additional integer link variables have an important consequence.
Using the Poisson summation formula
\eqn\psf{\sum_{k=-\infty }^{+\infty } {\rm e}^{i2\pi k z} =
\sum _{n=-\infty }^{+\infty } \delta (z-n) \ , }
we recognize that the sums over the integer link variables $\{m_{\mu }\}$
and $\{l_{\mu }\}$ enforce the following constraints:
\eqn\xdx{\eqalign{K_{\mu \nu } \left( B_{\nu }+{2\pi \over l} m_{\nu }
\right) &= {2\pi \over \kappa l^2} \ \beta _{\mu }\ ,\qquad
\beta _{\mu } \in Z\ , \cr
\hat K_{\mu \nu } \left( A_{\nu }+{2\pi \over l} l_{\nu}
\right) &= {2\pi \over \kappa l^2} \ \alpha _{\mu }\ ,\qquad
\alpha _{\mu } \in Z\ ,\cr }}
for all values of $A_{\mu }$, $B_{\mu }$, $m_{\mu }$ and $l_{\mu }$.
These have the immediate consequence of requiring a quantization
condition on the parameter $\kappa $:
\eqn\quc{\kappa = p \in Z.}
The integer variables $\alpha _{\mu }$ and $\beta _{\mu }$ are then
identified modulo $p$, which means $\alpha _{\mu }, \beta _{\mu } \in Z_p$.

Let us now consider the variation of the action \lam \ under a gauge
transformation $A_{\mu } \to A_{\mu }+d_{\mu } \Lambda $. For simplicity
let us take $\Lambda $ as a function of the first component $x^1$ only.
Under such a gauge transformation the lattice
action \lam \ changes by the surface term obtained by summing by
parts the second term. Since
the boundary conditions are such that the dual field strengths $F_{\mu }$
and $f_{\mu }$ vanish modulo $2\pi /l^2$ at infinity, we obtain
\eqn\cgt{\Delta S = \sum_{x^2, x^3} \ -ip \left[ \Lambda (x^1=+\infty ) \ n_+
- \Lambda (x^1=-\infty ) \ n_- \right] \ ,}
with $n_+$ and $n_-$ integers.
Gauge invariance requires that $\Delta S$ vanishes
modulo $i2\pi $. This is realized only if $\Lambda $ takes
the values $\Lambda = (2\pi /p) n$, $n\in Z_p$ at infinity.
Clearly, the same holds true for gauge transformations $B_{\mu } \to
B_{\mu }+d_{\mu }\Lambda $.
This means that both {\it global} gauge symmetries are actually broken down to
{\it discrete} $Z_p$ {\it symmetries}. An analogous phenomenon has been
encountered by Lee \kle \ in his investigation of continuum, compact
Chern-Simons theories.

In the following, we shall investigate how the coupling term affects the
topological excitations and their interactions. To this end
we decompose $n_{\mu }$ and $k_{\mu }$ as
\eqn\dec{\eqalign{n_{\mu } &\equiv lK_{\mu \nu } l_{\nu } +
a_{\nu }\ ,\cr
k_{\mu } &\equiv lK_{\mu \nu } m_{\nu } + b_{\nu }\ ,\cr }}
with $a_{\mu }$ and $b_{\mu }$ integers. The summations over
$\{ n_{\mu }\} $ and
$\{ k_{\mu }\} $ in \lam \ can then be traded for summations over
the new integers $\{a_{\mu }\}$ and
$\{b_{\mu }\}$. Accordingly, we can rewrite the partition
function in the following way:
\eqn\lan{\eqalign{Z &= \sum _{{\{a_{\mu }\}, \{l_{\mu }\} }\atop {
\{b_{\mu }\} , \{ m_{\mu }\} }} \int _{-\pi \over l}^{\pi \over l}
 {\cal D} A_{\mu } {\cal D} B_{\mu } \ {\rm exp}(-S) \ ,\cr
S &= \sum_{{\bf x}, \mu }\  {l^3\over 2e^2} \left[ K_{\mu \nu} \left(
A_\nu + {2\pi \over l}  l_\nu \right)  +{2\pi \over l^2} a_{\mu
} \right] ^2 - i{l^3 p \over 2\pi } \left( A_{\mu } + {2\pi \over
l} l_{\mu } \right) K_{\mu \nu } \left( B_{\nu } +{2\pi \over l}
m_{\mu } \right) \cr
&\ \ \ \ \ \ \ \ \ +{l^3\over 2g^2} \left[ K_{\mu \nu} \left(
B_\nu + {2\pi \over l}  m_\nu \right)  +{2\pi \over l^2} b_{\mu
} \right] ^2  \ ,\cr }}

At this point, we change variables,
\eqn\chv{\eqalign{A_{\mu } \ &\to \ A_{\mu }+ {2\pi \over l} \ l_{\mu } \ ,\cr
B_{\mu } \ &\to \ B_{\mu } + {2\pi \over l} \ m_{\mu } \ ,\cr }}
in the integrations over the gauge fields. The sums over the integers
$\{l_{\mu }\}$ and $\{m_{\mu }\}$ can now be carried out explicitly, with
the effect of extending the integration interval for the gauge fields
from $[-\pi /l, +\pi /l]$ to $(-\infty, +\infty )$:
\eqn\its{\eqalign{Z &= \sum_{{\{a_{\mu }\}} \atop {\{b{\mu }\}}}
\int _{-\infty }^{+\infty } {\cal D} A_{\mu } {\cal D} B_{\mu } \
{\rm exp}(-S)\ ,\cr
S &= \sum_{{\bf x}, \mu }\  {l^3\over 2e^2} F_{\mu }^2 -i{l^3 p
\over 2\pi } A_{\mu }K_{\mu \nu }B_{\nu }+{l^3\over
2g^2} f_{\mu }^2  \cr
&\ \ \ \ \ \ \ \ \ +{2\pi ^2\over le^2}\  a_{\mu }^2
+{2\pi ^2\over lg^2}\  b_{\mu }^2
+ {2\pi l\over e^2} A_{\mu } \hat K_{\mu \nu } a_{\nu } + {2\pi l
\over g^2 } B_\mu \hat K_{\mu \nu} b_\nu\ . \cr }}

In a last step, we carry out the Gaussian integrations over $A_{\mu }$
and $B_{\mu }$. To this end, we introduce the usual gauge fixing terms;
these, however, drop out from the final answer, since the gauge fields
are coupled to topologically conserved currents $\hat K_{\mu \nu }a_{\nu }$
and $\hat K_{\mu \nu }b_{\nu }$.
The result of the Gaussian integrations takes the form
$Z=Z_0\cdot Z_{\rm Top}$, where $Z_0$ is the lattice partition function for the
non-compact, Euclidean version of the model,
\eqn\ncv{Z_0=\int _{-\infty }^{+\infty } {\cal D}A_{\mu }{\cal D}B_{\mu }
\ {\rm exp} \sum_{\bf x, \mu } \left\{ -{l^3\over 2e^2} F_{\mu }^2 +
{il^3p\over 2\pi }A_{\mu }K_{\mu \nu }B_{\nu }-{l^3\over 2g^2} f_{\mu }^2
\right\} \ ,}
and $Z_{\rm Top}$ is given by
\eqn\top{\eqalign{Z_{\rm Top} &=\sum_{{\{a_{\mu }\}}
\atop {\{b_{\mu }\}}} {\rm exp } \left( - S_{\rm Top} \right) \cr
S_{\rm Top} &= \sum_{{\bf
x}, \mu } -{e^2\over 2l^3} \ J_{\mu } {\delta _{\mu \nu} \over {m^2-\nabla ^2}}
J_{\nu } -{g^2\over 2l^3} \ K_{\mu }
{\delta _{\mu \nu } \over {m^2-\nabla ^2}} K_{\nu } \cr
&\ \ \ \ \ \ \ \ -i{e^2g^2p\over 2\pi l^3} \ J_{\mu } {K_{\mu \nu}\over
{\nabla ^2 (m^2-\nabla ^2)}} K_{\nu } +{2\pi ^2\over le^2} a_{\mu }^2
+{2\pi ^2\over lg^2} b_{\mu }^2 \ ,\cr }}
with the currents $J_{\mu }$ and $K_{\mu }$ defined by
\eqn\nwc{\eqalign{J_{\mu } &\equiv {2\pi l\over e^2} \ \hat K_{\mu \nu }
a_{\nu } \ ,\cr
K_{\mu } &\equiv {2\pi l\over g^2} \ \hat K_{\mu \nu } b_{\nu } \ ,\cr }}
and the mass $m$ given in \mas .
In terms of the integer variables $a_{\mu }$ and $b_{\mu }$, the action
takes its final form
\eqn\tup{\eqalign{S_{\rm Top} = \sum_{{\bf
x}, \mu } \ &{2\pi ^2\over le^2} \  a_{\mu } {{m^2 \delta _{\mu \nu
} -d_{\mu }\hat d_{\nu }} \over {m^2-\nabla ^2}}a_{\nu }
+{2\pi ^2 \over lg^2}\  b_{\mu }{{m^2
\delta _{\mu \nu }-d_{\mu }\hat d_{\nu }}\over {m^2-\nabla ^2}}
b_{\nu } \cr
&+i{2\pi p \over l}\  a_{\mu }
{K_{\mu \nu }\over {m^2-\nabla ^2}}
b_{\nu }\ .\cr}}

The partition function $Z_{\rm Top}$ represents the contribution
of the {\it topological excitations} $a_{\mu }$
and $b_{\mu }$, due to the {\it compactness} of the two gauge
symmetries. The string-like excitations $a_{\mu }$ and $b_{\mu }$
originate as the integer parts of $F_{\mu }$ and $f_{\mu }$ respectively,
and have therefore the obvious interpretation of {\it magnetic flux strings}.
With the interpretation
\dcu , however, $b_{\mu }$ represents {\it charge current strings}.

The strings can be closed (rings), in which case $\hat d_{\mu }a_{\mu }=0$
and $\hat d_{\mu }b_{\mu }=0$, or open, in which case they terminate on
monopole-antimonopole pairs. In our Euclidean formalism, these monopoles
describe tunneling events corresponding to the creation or destruction
of $p$ localized, elementary fluxes or charges. Fluxes and charges are
indeed conserved only modulo $p$, due to the discrete gauge symmetries
$Z_p$. Note that, in our context, charge does not refer to the particles
of the underlying microscopic model; rather it resides on localized,
collective {\it quasi-particle excitations}.

The mechanism leading to string-like topological excitations has its
origin in the mixed Chern-Simons coupling \foot{The same mechanism is
responsible for confinement of monopoles in Maxwell-Chern-Simons theory
\dst .}. For $p=0$ (and therefore $m=0$), $S_{\rm Top}$ reduces to a sum
of uncoupled Coulomb gases of monopoles, as expected. Suppose now we start
with an isolated monopole for one of the gauge fields: when we turn on the
coupling, its otherwise unobservable Dirac string acquires "electric
charge" coupled to the other gauge field and becomes thus an observable,
physical entity. Moreover, as is evident from \tup , this charge endows
the string with a finite energy per unit length: therefore, infinite
open strings do not contribute to the partition function and only
closed or finite open strings survive. This means that the mixed
Chern-Simons coupling effectively confines the monopoles.

The self-duality of the original model \dla \ is reflected in the invariance
of $S_{\rm Top}$ under the duality transformation
\eqn\lad{\eqalign{a_{\mu } \ &\longleftrightarrow \ b_{\mu } \ ,\cr
e\ &\longleftrightarrow \ g \ . \cr }}
Actually, on the lattice, self-duality is only an approximate symmetry
due to the imaginary last term in \tup , which contains the lattice
Chern-Simons operator $K_{\mu \nu }$. Under the above duality
transformation we obtain an action identical to \tup , with the only
difference that the last term is written
in terms of $\hat K_{\mu \nu }$ instead of $K_{\mu \nu }$. Note however,
that the theories defined with $K_{\mu \nu }$ and $\hat K_{\mu \nu }$
are completely equivalent.

\subsec{Wilson and t'Hooft loops}
In order to distinguish the various possible phases of the model we
introduce two order parameters, namely the Wilson loop operators
\pol \ for the two gauge fields. Since the gauge field $B_{\mu }$
couples to the magnetic flux, the corresponding Wilson loop operator
coincides with the magnetic order parameter first introduced by
t'Hooft \ref\oft{G. t'Hooft, Nucl. Phys. B138 (1978) 1.} \ \eli .
We shall call it the t'Hooft loop operator.
The vacuum expectation values of the Wilson and t'Hooft loop operators
determine the interaction potential between external test charges and
fluxes \pol , and provide thus a criterion for confinement.

The lattice version of these operators is given by
\eqn\wtl{\eqalign{L_W &= {\rm exp} \  iq\sum_{{\bf x}, \mu}
lq_{\mu} A_{\mu } \ ,\cr
L_H &= {\rm exp } \ i\phi \sum_{{\bf x}, \mu } l\phi _{\mu }
B_{\mu } \ ,\cr }}
where the integers $q$ and $\phi $ represent the strengths of the
external test charge and flux, respectively, and $q_{\mu }$ and
$\phi _{\mu }$ vanish everywhere but on the links of the loops, where
they take the value 1. Since the loops are closed, they satisfy
\eqn\cll{\hat d_{\mu } q_{\mu } = \hat d_{\mu } \phi _{\mu } =0 \ .}

We shall concentrate exclusively on the models with the matter fields
carrying a charge $p>1$. Correspondingly, we choose $q, \phi < p$ in our
order parameters. It is in fact to be expected that Wilson loops fail
as a criterion for confinement for $p=1$, as in the Abelian Higgs
model \frash .

The computation of the expectation value of the order parameters
implies the evaluation of the following integral:
\eqn\iwh{\eqalign{L &\equiv \langle {\rm exp } \sum_{{\bf x}, \mu }
\left( iq lq_{\mu } A_{\mu } + i\phi l\phi _{\mu }B_{\mu } \right)
\rangle \cr
&= {1\over Z} \ \sum_{{\{n_{\mu }\}, \{l_{\mu }\}}
\atop {\{k_{\mu }\}, \{m_{\mu }\}}}
\int _{-{\pi \over l}}^{+{\pi \over l}} \ {\cal D}A_{\mu } {\cal D}
B_{\mu } \ {\rm exp } \left\{ -S + \sum_{{\bf x}, \mu } \left( iq
lq_{\mu }A_{\mu } + i\phi l\phi _{\mu }B_{\mu } \right) \right\} \ ,\cr }}
with $S$ given in \lam . Following exactly the same steps as in the
evaluation of the partition function, we obtain
\eqn\ist{L= {1\over Z_{\rm Top}} \ \sum_{\{a_{\mu }\}\atop \{b_{\mu }\}}
{\rm exp }(-W) \ ,}
where $W=S_{\rm Top}$ with the currents \nwc \ redefined as
\eqn\rec{\eqalign{J_{\mu } \ &\to \ J_{\mu } + iqlq_{\mu } \ ,\cr
K_{\mu } \ &\to \ K_{\mu } + i\phi l \phi _{\mu } \ .\cr }}
This leads to the following representation of the order parameter:
\eqn\fop{\eqalign{L &= {\rm exp} \left( -W_0 \right) \ {1\over Z_{\rm Top}}\
\sum _{\{a_{\mu }\}\atop \{b_{\mu }\}} {\rm exp }\left( -S_{\rm Top} -
W_{\rm Top} \right) \ ,\cr
W_0 &= \sum _{{\bf x}, \mu } {q^2 e^2 \over 2l} \ q_{\mu }
{\delta _{\mu \nu }\over {m^2-\nabla ^2}} q_{\nu } + {\phi ^2 g^2\over
2l} \ \phi _{\mu } {\delta _{\mu \nu }\over {m^2-\nabla ^2}} \phi_{\nu} \cr
&\ \ \ \ \ \ \ \ \ -i{2\pi q\phi m^2\over pl} \ q_{\mu }
{K_{\mu \nu }\over{\nabla ^2\left(m^2-\nabla ^2\right)}} \phi _{\nu} \ ,\cr
W_{\rm Top} &= \sum_{{\bf x}, \mu } -i{2\pi q\over l} \ q_{\mu }
{\hat K_{\mu \nu } \over {m^2-\nabla ^2}} a_{\nu } -i {2\pi \phi \over l}
\ \phi _{\mu }{\hat K_{\mu \nu } \over {m^2-\nabla ^2}} b_{\nu } \cr
&\ \ \ \ \ \ \ \ - qe^2p \ A^q_{\mu } {K_{\mu \nu }
 \over {m^2-\nabla ^2}} b_{\nu }
- \phi g^2 p\ A^{\phi }_{\mu }
{\hat K_{\mu \nu } \over
{m^2-\nabla ^2}} a_{\nu } \ ,\cr }}
where we have introduced the elementary area elements $A^q_{\mu }$ and
$A^{\phi }_{\mu }$ of the closed loops:
\eqn\are{\eqalign{q_{\mu } &= l\hat K_{\mu \nu } \ A_{\nu }^q \ ,\cr
\phi _{\mu } &= l\hat K_{\mu \nu} \ A_{\nu }^{\phi } \ .\cr }}
These vanish everywhere but on
the links perpendicular to the elementary plaquettes spanning the
minimal area enclosed by the loops, where they take the value 1.

In \fop , the factor ${\rm exp}(-W_0)$ represents the contribution from
the massive, propagating modes described by the partition function
$Z_0$ in \ncv . The second factor, instead, describes the
contribution from the topological excitations.

The first two terms in $W_0$ describe the screened Coulomb interaction
mediated by the massive gauge particles. The third term, instead, represents
the {\it Aharonov-Bohm interaction} between charges $q$ and fluxes $\phi $
separated on distances much larger $1/m$. In this case one can in fact
neglect all the off-diagonal terms in the kernel $1/(m^2-\nabla ^2)$ and
retain only its diagonal term $1/m^2$, thereby obtaining the expression
\eqn\abi{\sum_{{\bf x}, \mu } \ q_{\mu } {lK_{\mu \nu} \over {l^2\nabla ^2}}
\phi _{\nu }\ .}
This is an {\it integer}, as can be easily recognized by inserting the
representation \are \ for $\phi _{\mu }$. This
integer is a lattice version of the {\it Gauss linking number}
of the two closed loops. The Aharanov-Bohm interaction
vanishes thus for $q\phi =np$, $n\in Z$, which is the celebrated
Dirac quantization condition (in our units) for a discrete $Z_p$ gauge
theory.

The contribution $W_{\rm Top}$ vanishes when both types of topological
excitations are suppressed, i.e. when the partition function $Z_{\rm Top}$
is dominated by the saddle point $a_{\mu }=0$, $b_{\mu }=0$.
In this case, nothing is changed with respect to the above picture.
Charges and fluxes come in neutral bound states with binding energies
of order ${\rm log} (1/ml)$ and are essentially free for $ml > O(1)$.
The photon has a {\it topological mass} due to the Chern-Simons mechanism.
We shall therefore call this phase of the theory the {\it Chern-Simons
phase}. For $ml > O(1)$ it can be identified with the mixed phase of a
type-II superconductor. This Chern-Simons phase corresponds to the
Coulomb phase of (3+1)-dimensional $Z_N$ gauge models. In (2+1) dimensions
a pure Coulomb phase does not exist since the photon is always massive
due to the Chern-Simons mechanism.
It is this phase of the model which was investigated in \mav .

This picture can be drastically changed when one type of topological
excitations condenses.
To see this let us suppose that $Z_{\rm Top}$
is dominated by the saddle point $b_{\mu }=0$, while the formation
of long $a_{\mu }$ strings is favoured. There are two types of strings:
open ones and closed ones. For the planar quantum system the former
describe tunneling events corresponding to the formation and subsequent
destruction of localized fluxes $p$; the latter describe instead the
formation and annihilation of neutral pairs of such fluxes.
The condensation of long strings in the three-dimensional statistical
mechanics problem indicates that the ground state of the planar quantum
system consists of a magnetic condensate.
The total flux number of the ground state fluctuates around zero due to
the monopoles at the end of the open strings.
We now show that such fluctuations confine electric charges.

First of all let us remark that a condensation of long $a_{\mu }$
strings is accompanied by a condensation of closed strings
$\hat K_{\mu \nu }a_{\nu }$, representing circular electric currents.
These couple to external test fluxes $\phi $ via the last term in
$W_{\rm Top}$. However, these electric currents form {\it tiny} loops
around the long magnetic flux lines $a_{\mu }$  and their effects on test
fluxes are essentially negligible \pol \ (apart from a
renormalization of the coupling constant of the screened Coulomb potential).

The only relevant term in $W_{\rm Top}$ is therefore the first term,
coupling $a_{\mu }$ to the Wilson loop $q_{\mu }$. For $ml \gg 1$, we can
neglect $\nabla ^2$ in the interaction kernel $1/\left( m^2 -\nabla ^2
\right)$.
In this case, the coupling reduces to a contact term between the magnetic
flux string $a_{\mu }$ and the tiny flux rings $K_{\mu \nu }q_{\nu }$
encircling the Wilson loop $q_{\mu }$. This contact term will contribute
only a perimeter law to the Wilson loop expectation value and essentially
renormalizes the first term in $W_0$. The same argument can be repeated
when the electric strings $b_{\mu }$ condense: we thus conclude that for
$ml \gg 1$ the system possesses only the Chern-Simons phase.

Let us now concentrate on the case $ml \ll 1$ and let us consider Wilson
loops with typical dimension $L$ in the range $l \ll L \ll 1/m$. For
such loops we can neglect $m^2$ in the interaction kernel with all
strings $a_{\mu }$ (first term in $W_{\rm Top}$) passing through the
surface spanned by the loop. As a consequence, these strings become
unobservable for the Wilson loop, which couples only to the monopoles
at their end:
\eqn\dtrs{\sum_{{\bf x}, \mu } i{2\pi q\over l} \ q_{\mu }
{\hat K_{\mu \nu } \over \nabla ^2} a_{\nu } =
\sum_{\bf x} -i {2\pi q\over l} \ \hat d_{\mu }A^q_{\mu } {1\over \nabla ^2}
Q \ ,}
since $A^q_{\mu }a_{\mu }$ is an integer. Here $Q\equiv l\hat d_{\mu }
a_{\mu }$ represents the monopoles.
This means that only the longitudinal degrees of freedom of $a_{\mu }$
couple to the Wilson loop. As it is easy to see by inserting the
representation
\eqn\repr{a_{\mu } = ld_{\mu }\omega \ , \qquad \qquad
l^2\nabla ^2 \omega = Q \ ,}
into $S_{\rm Top }$, these describe a Coulomb gas of magnetic monopoles.
Charges are therefore confined by the familiar Polyakov
mechanism \pol .

Actually, the above computation indicates only the presence of a linear
potential between charges up to scales $1/m$. However, since the string
tension is of order $e^2/l$ \pol , the energy required to separate two
charges is at least of order $e^2/(ml) = (2\pi e / pg) (1/l)$. In the
next section we shall show that the condensation of magnetic strings
is favoured for large values of $e/g$. The binding energy is therefore
much larger than the ultraviolet cutoff $\Lambda =1/l$ and charges are
effectively confined.

We thus conclude that the phase in which the condensation of magnetic
strings $a_{\mu }$ is favoured is a {\it confinement (or insulating) phase}.
The same arguments repeated for strings $b_{\mu }$ lead to the
corresponding conclusion that the phase in which the condensation of
electric strings $b_{\mu }$ is favoured is a {\it Higgs (or superconducting)
phase}.

\subsec{Phase structure analysis}
As explained above, in order to establish the phase structure of the
model as a function of its parameters, we need to analyze the conditions
for the condensation of the topological excitations. To this end, we
shall use the same free energy arguments adopted in the analysis of the
related (3+1)-dimensional models \eli \cra . In these arguments, the
condition for condensation of strings is established by analyzing the
balance between the self-energy of a string and its entropy.

The free energy of a string of length $L=lN$ carrying
magnetic and electric quantum numbers $a$ and $b$ is essentially
\eqn\pse{\beta F= \left( {2\pi ^2 \over le^2} (ml)^2 G(ml)\ a^2 +
{2\pi ^2 \over lg^2} (ml)^2 G(ml)
\ b^2 -\gamma \right) N \ ,}
where $G(ml)$ is the diagonal element of
the lattice kernel $G({\bf x}-{\bf y})$
representing the inverse of the operator $l^2 (m^2-\nabla ^2)$.
Clearly, this diagonal element also depends on the dimensionless
parameter $ml$.
The last term in \pse \ represents the entropy
of the string: the parameter $\gamma $ is given roughly by $\gamma ={\rm ln}
5$,
since at each step the string can choose between 5 different directions.
In a dilute instanton approximation, in which all values $a_{\mu },
b_{\mu } \ge 2$ are neglected, it can be proved that the correct value
of $\gamma $ is the same for open and closed strings \ref\eis{M. B.
Einhorn and R. Savit, Phys. Rev. D19 (1979) 1198.} .
In \pse \ we
have neglected all subdominant functions of $N$, like a ${\rm ln} N$
correction to the entropy and a constant term due to the monopole
contribution to the energy for open strings. Moreover, we have neglected
the imaginary term in the action \tup . This can be justified
self-consistently, since the contribution of this term vanishes in
all phases of the model, as we now show.

The condition for the condensation of topological excitations is
obtained by minimizing the free energy \pse \ as a function of $N$.
If the coefficient of $N$ in \pse \ is positive, the minimum of
$\beta F$ is obtained for $N=0$ and topological excitations are
suppressed. If instead the same coefficient is negative, the minimum
of $\beta F$ is obtained for $N= \infty $ and the system will favour
the formation of long strings. Topological excitations with quantum
numbers $a$ and $b$ condense therefore if
\eqn\coc{{2\pi ^2 \over le^2 \delta }\ a^2 + {2\pi ^2 \over lg^2 \delta } \ b^2
<1\ ,}
where we have introduced
\eqn\nepa{\delta \equiv {\gamma \over {(ml)^2 G(ml)}}\ .}
This new parameter is clearly also a function of $ml$.
When two or more condensates are possible, one has to choose the one
with the lowest free energy.

This condensation condition describes the interior of an ellipse
with semi-axes $le^2\delta /2\pi ^2$ and $lg^2\delta /2\pi ^2$ on a square
lattice of integer magnetic and electric charges. The phase diagram
is obtained by investigating which points of the integer lattice lie
inside the ellipse as its semi-axes are varied. We find it convenient
to present the result in terms of the parameters $lm$ and $e/g$.
For $lm \gg 1$ we have only the Chern-Simons phase, for all values of
$e/g$. For $ml \ll 1$ we obtain instead the following phase structure:
\eqn\pst{\eqalign{{\delta lm\over \pi p} >1 &\to \cases{{e\over g}<1\ ,
& Higgs (superconducting)\ ,\cr
{e\over g} >1\ , & confinement (insulating)\ ,\cr} \cr
{\delta lm\over \pi p}<1 &\to \cases{{e\over g} < {\delta lm\over \pi p} \ ,
& Higgs (superconducting)\ ,\cr
{\delta lm\over \pi p}< {e\over g} < {\pi p\over \delta lm} \ , & Chern-Simons\
,\cr
{e\over g}>{\pi p\over \delta lm} \ , & confinement (insulating) \ .\cr } \cr
}}
As expected, the phase diagram is symmetric around the self-dual point
$e/g =1$. For small $e/g$ we obtain a Higgs (superconducting) phase, while
for large $e/g$ the model is in a confinement (insulating) phase. However,
for $\delta lm/\pi p<1$, these phases do not extend all the way to $e/g=1$;
rather, an intermediate Chern-Simons phase opens
up between the Higgs and confinement phases.
The results \pst \ were derived assuming $ml \ll 1$.
The presence or absence of an intermediate Chern-Simons phase
depends therefore on the exact form of the function
$lm \delta (lm)$ for $lm \ll 1$.

We conclude this section by stressing that an
insulating-superconducting quantum phase transition is actually
observed experimentally \ref\moi{L. J. Geerligs, M. Peters, L. E. M. de Groot,
A. Verbruggen and J. E. Mooij, Phys. Rev. Lett. 63 (1989) 326; R. Fazio,
A. van Otterlo, G. Sch\"on, H. S. J. van der Zant and J. E. Mooij, Helv.
Phys. Acta 65 (1992) 228.} in planar Josephson junction arrays at
extremely low temperatures. This further confirms that the self-dual
lattice gauge theory \lam \ (for $\kappa =2$) captures the essential
physics of planar Josephson junction arrays and raises the question
wether the intermediate Chern-Simons phase might be experimentally
accessible at even lower temperatures.

\subsec{Including (3+1)-dimensional effects}
Up to now we have discussed only purely planar effects. As we pointed
out in section 2, however, the photon kinetic term has to be modified
as in \mpr \ in order to describe (3+1)-dimensional electromagnetism
coupled to planar matter. In the following we investigate how this
modification affects the phase structure of the model.

The new lattice model is easily obtained by substituting the first
term in the action \lam \ with
\eqn\npt{{l^3\over 2e^2} \ \left(F_{\mu }+{2\pi \over l^2}n_{\mu }\right)
{1\over \sqrt{-\nabla ^2}} \left(F_{\mu }+{2\pi \over l^2}n_{\mu }
\right) \ ,}
where $e^2$ is now dimensionless.
All the steps leading to \tup \ and \fop \ can be exactly repeated.
The resulting modified expressions for $S_{\rm Top}$, $W_0$ and
$W_{\rm Top}$ are given by:
\eqn\mew{\eqalign{S_{\rm Top} &= \sum_{{\bf x}, \mu } \ {2\pi ^2 \over le^2}
\ a_{\mu } {{\mu \sqrt{-\nabla ^2} \delta _{\mu \nu } - d_{\mu }\hat d_{\nu }}
\over {-\nabla ^2 \left( \sqrt{-\nabla ^2} + \mu \right) }} a_{\nu }
+{2\pi ^2\over lg^2}\ b_{\mu } {{\mu \sqrt{-\nabla ^2} \delta _{\mu \nu}
-d_{\mu }\hat d_{\nu }} \over {\sqrt{-\nabla ^2} \left( \sqrt{-\nabla ^2}
+\mu \right) }} b_{\nu } \cr
&\ \ \ \ \ \ \ \ \ +i{2\pi p\over l} \ a_{\mu } {K_{\mu \nu }\over
{\sqrt{-\nabla ^2} \left( \sqrt{-\nabla ^2}+\mu \right) }} b_{\nu } \ ,\cr
W_0 &= \sum_{{\bf x}, \mu } \ {q^2 e^2\over 2l} \ q_{\mu } {\delta _{\mu \nu }
\over {\sqrt{-\nabla ^2} +\mu }} q_{\nu } + {\phi ^2 g^2 \over 2l}
\ \phi _{\mu } {\delta _{\mu \nu } \over {\sqrt{-\nabla ^2} \left(
\sqrt{-\nabla ^2} +\mu \right) }} \phi _{\nu } \cr
&\ \ \ \ \ \ \ \ \ -i{2\pi q\phi \mu \over pl} \ q_{\mu } {K_{\mu \nu } \over
{\nabla ^2 \left( \sqrt{-\nabla ^2} +\mu \right) }} \phi _{\nu } \ ,\cr
W_{\rm Top} &= \sum _{{\bf x}, \mu } \ -i{2\pi q\over l} \ q_{\mu }
{\hat K_{\mu \nu } \over {\sqrt{-\nabla ^2} \left( \sqrt{-\nabla ^2} + \mu
\right) }} a_{\nu } -i {2\pi \phi \over l} \ \phi_{\mu } {\hat K_{\mu \nu }
\over {\sqrt{-\nabla ^2} \left( \sqrt{-\nabla ^2} +\mu \right) }} b_{\nu } \cr
&\ \ \ \ \ \ \ \ \ -qe^2 p  \ A^q_{\mu } {K_{\mu \nu }
\over {\sqrt{-\nabla ^2} +\mu  }} b_{\nu }
-\phi g^2 p\ A^{\phi }_{\mu }
{\hat K_{\mu \nu }\over
{\sqrt{-\nabla ^2} \left( \sqrt{-\nabla ^2 } + \mu  \right) }}
a_{\nu } \ , \cr }}
where the mass $\mu $ is given in \mma .

In this model, the Chern-Simons phase (characterized by the absence
of topological excitations) consists of charges interacting via
a screened $1/r$-interaction (Coulomb interaction in (3+1) dimensions)
and fluxes interacting via a logarithmic potential (Coulomb interaction
in (2+1) dimensions) up to scales $1/\mu $ and a $1/r$-interaction
on larger scales. The photon is still massive and it is this photon
mass which screens the 2- and 3-dimensional Coulomb interactions on
scales $1/\mu $. Note that the Aharonov-Bohm interaction for charges
and fluxes separated by distances much larger than $1/\mu $ is
unaffected by the modification \npt .

Let us now consider again the effects of the condensation of topological
excitations. Suppose first that the electric strings $b_{\mu }$
condense. In this case one can repeat verbatim the analysis of the
preceding section with the same conclusion that this is
a Higgs (superconducting) phase. When the magnetic strings $a_{\mu }$
condense one can also repeat the above analysis; however in this case
there is a crucial difference with respect to the purely planar case.
The Wilson loop still couples only to the longitudinal degrees of
freedom of $a_{\mu }$, which are represented by the magnetic monopoles.
However, using the representation \repr \ in the first term of
$S_{\rm Top}$ we obtain
\eqn\mgmo{S_Q= \sum_{\bf x} {2\pi ^2\over e^2 l^3} \ Q
{1\over {-\nabla ^2 \sqrt{-\nabla ^2}}} Q \ .}
This is the Hamiltonian for magnetic monopoles with a logarithmic
interaction (at large distances) in 3 dimensions. The logarithmic
interaction is confining. The same free energy arguments used to derive
the phase structure of a two-dimensional Coulomb gas suggest the existence
of a strong coupling phase at low $e^2$, in which the monopoles are
confined. We expect this phase to be realized for the small value of the
fine structure constant $e^2/4\pi $. Since monopoles are confined, they
cannot screen the dipole sheet $\hat d_{\mu }A^q_{\mu }$ in \dtrs \ and
the Wilson loop does not acquire an area law. This means that for a
sufficiently weak Coulomb interaction of charges in the model \mpr \ there
is no confinement phase, which is the expected result.

The transition point between the Chern-Simons phase and the Higgs
(superconducting) phase is determined by the condition for condensation
of the electric strings $b_{\mu }$. In analogy to \coc \ this is given by
\eqn\comp{{p^2 e^2 \over {2\delta (\mu l)}} \ b^2 < 1\ .}
Here $\delta (\mu l) \equiv \gamma /G(\mu l)$, and $G(\mu l)$ is the
diagonal element of the lattice kernel representing the inverse of
$l(\sqrt{-\nabla ^2} +\mu )$. We thus obtain the following phase structure:
\eqn\phmp{\eqalign{ \delta (\mu l) &> {p^2 e^2 \over 2}\ , \qquad \qquad
{\rm Higgs\ (superconducting)} \ ,\cr
\delta (\mu l) &< {p^2 e^2 \over 2}\ , \qquad \qquad
{\rm Chern-Simons} \ .\cr }}
The function $\delta (\mu l)$ has the following asymptotic
behaviour,
\eqn\asbe{\eqalign{\mu l \ &\to \ 0\ , \qquad \qquad \delta (\mu l)
= {\rm const.}\ ,\cr
\mu l \ &\to \infty \ , \qquad \qquad \delta (\mu l) \propto \mu l \ .\cr }}
This implies that the system is always in the Higgs (superconducting) phase
for a sufficiently large mass gap $\mu $ (pairing gap in the underlying
microscopic model \mav ). The possible presence of the additional Chern-Simons
phase at $T=0$ depends again on the details of the function $\delta (\mu l)$.

\newsec{Oblique confining model: non-perturbative analysis}
In this section we shall consider the models
with a topological Chern-Simons term for the matter gauge field $B_{\mu }$.

Formulating a compact lattice version of \nfo \ and \rqh \ requires
again the introduction of integer link variables, which
enforce constraints analogous to \xdx . These require the quantization
of both parameters $\kappa $ and $\eta $:
\eqn\quan{\kappa = p\in Z\ ,\qquad \qquad \eta =n \in Z\ .}
The ensuing gauge group depends crucially on the commensurability of
$p$ and $n$. If $p$ and $n$ are coprime, the original $U(1)\times U(1)$
global gauge group is completely broken. If, instead, $p$ and $n$ have
a (maximal) common factor $r$, the residual discrete gauge symmetry
is $Z_r\times Z_r$.

Given the representations \nfo \ and \rqh , we can just make use of
the results of the previous section, provided we make the following
substitutions:
\eqn\sus{\eqalign{g \ &\to \ g'\ ,\cr
m \ &\to \ M\ ,\cr
\mu \ &\to \ \Delta \ ,\cr
a_{\mu } \ &\to \ \left( a_{\mu } +{n\over p} b_{\mu } \right) \ ,\cr
\phi \ &\to \ \left( \phi + {n\over p} q \right) \ .\cr }}
We still have a Chern-Simons phase and, for the purely planar model,
a confinement phase; however the
third possible phase changes completely its character. In the model of
the previous section, this third phase was characterized by the
condensation of electric strings $b_{\mu }$, while $a_{\mu }=0$.
With the above substitutions, this means a condensation of strings
carrying both {\it electric and magnetic quantum numbers a and b} in the
ratio $a/b=-n/p$, i.e. of dyonic strings. Correspondingly, this condensation
implies that particles with quantum numbers $\phi + (n/p)q \ne 0$ are
confined. The ground state of the planar quantum system consists
of a dyonic condensate; excitations carry both electric charge and magnetic
flux in the same ratio as in the condensate, i.e. $\phi + (n/p)q =0$.

If $p=rP$ and $n=rN$, with $P$ and $N$ coprime, the elementary excitations
carry fractional (in units of the fundamental charge $p$) charge $P/p$
and magnetic flux $N$. These excitations are {\it anyons} \ref\wil{For a
review see F. Wilczek, "Fractional Statistics and Anyon Superconductivity",
World Scientific, Singapore (1990).} with fractional statistics
$PN/p$ (modulo 2) originating in the Aharonov-Bohm interaction of an
elementary charge $P$ with the flux $N$ carried by a second elementary charge.

This phase of the model is an {\it oblique confinement} phase \hoo \ \cra .
It is characterized by a gap, the absence of longitudinal conductivity and
the presence of a quantized Hall conductivity. We thus identify the
corresponding ground state of the system as Laughlin's {\it incompressible
quantum fluid} \ref\lau{R. B. Laughlin, "The Incompressible Quantum Fluid",
in \gip .}.

We expect this phase to be realized when all offset charges (external
magnetic fluxes for the model in (3+1) dimensions) can be attached to the
vortices (charges for the model in (3+1) dimensions), i.e. for $n=p$. In
this case, the fractional charge and statistics of elementary excitations
are both given by $1/p$.

In \hac \ we have derived the Hall current characterizing the oblique
confinement phase (for $n=p$) as
\eqn\rhac{j^i_H = {(p/2)^2 \over 2\pi } \ {1\over (p/2)} \ \epsilon ^{ij}
E^j \ .}
For the smallest allowed value of $p$, the matter gauge fields describe
therefore particles of charge $1/2$.  If we don't want to describe
physical electrons as Cooper pairs of charge $1/2$ particles we must
rescale all charges by a factor 2 and therefore
$p\to 2p$, $n \to 2n$.
Moreover, $p$ has to be odd if electrons are to retain
their fermionic character. In the resulting description, the physical
electron is identified with a particle of charge $p$: all charges $1 \dots
p$ represent fractional charge particles. As a consequence of the original
compact gauge symmetry, all charges are quantized in integer units:
fractional Hall states are thus described by increasing the charge of
the electron.

Correspondingly, we describe bosonic Cooper pairs by choosing $p$ even
(after the above rescaling).
This is the relevant situation for applications
to Josephson junction arrays. As we showed in \eff , by integrating out
the vortex degrees of freedom one obtains an effective theory for the
charges which does not contain any Chern-Simons term. Conversely, if we
integrate out the charge degrees of freedom we obtain an effective action
for the vortices which contains a Chern-Simons term with coefficient
$p/4\pi $ (after the above rescaling). Repeating the computation leading
to \hac , we recognize that the oblique confinement phase is a quantum
Hall regime for the vortices. Had we added originally a Chern-Simons term
for the $A_{\mu }$ gauge field, we would obtain correspondingly a quantum
Hall regime for the charges. In this case, the Hall conductivities would
take the form
\eqn\pohc{\sigma _H = {(2e)^2 \over 2\pi } \ {1\over p} \ , \qquad
\qquad p={\rm even} \ .}

In the purely planar model, the oblique confinement phase described above
is realized for small $e/g'$ ($E_C/E_J$). In this context, ${g'}^2$ has to
be understood as the physical, renormalized (by the external offset charges)
Josephson coupling. As we explained above, this corresponds to a quantum
Hall regime for the vortices. Above a critical value for $E_C/E_J$, the
system undergoes a transition to a confinement (insulating) phase. Again, the
presence of a possible intermediate Chern-Simons phase depends on the detailed
behaviour of the function $lM \delta (Ml)$. In presence of external magnetic
fields, correspondingly, we obtain a quantum Hall regime for charges for
large values of $e'/g$ ($E_C/E_J$). In this case, the system is in a
Higgs (superconducting) phase for small values of the same parameter.

In the model including (3+1)-dimensional effects, we obtain a flux unbinding
transition from the oblique confinement phase (quantum Hall regime) to a
Chern-Simons phase when the gap $\Delta $ is lowered below a critical value.
Due to the $1/r$ interactions among charges in this latter phase, it is
to be expected that this phase immediately cristallizes at the low
temperatures in which quantum Hall experiments are performed. Note that the
critical gap is an increasing function of $p$, as is evident from \phmp .
This explains the lesser stability of quantum Hall states with smaller
filling fraction.

It is known \ref\gmr{S. M. Girvin and A. H. MacDonald, Phys.
Rev. Lett. 58 (1987) 1252; N. Read, Phys. Rev. Lett. 62 (1989) 86.}
\ that the microscopic Laughlin wave functions \lau for
the incompressible quantum fluids can be viewed as quantum states
in which an odd number of statistical fluxes are bound to
the electrons. This fact is at the basis of Jain's theory
\ref\jai{For a review see: J. K. Jain, Adv. in Phys. 41 (1992) 1.}\
of composite electrons and of most field theoretic treatments
\ref\flo{S. C. Zhang, T. Hansson and S. Kivelson, Phys. Rev.
Lett. 62 (1989) 82; A. Lopez and E. Fradkin, Phys. Rev. B44
(1991) 5246, Phys. Rev. Lett. 69 (1992) 2126, Nucl. Phys. B
(Proc. Supp.) 33C (1993) 67.}\ of the quantum Hall effect.
Our results demonstrate how the key aspect of the
quantum Hall effect is oblique confinement by the Polyakov monopole
mechanism. Given the non-perturbative nature of our treatment we
could explicitly derive the existence of a {\it critical gap} for the formation
of incompressible quantum fluids. We believe that this is a new and
important result in the framework of effective field theories for the
quantum Hall effect.

\newsec{Concluding remarks}
We would like to conclude this paper with the following two observations.
First, our results suggest that Josephson junction arrays might provide
an easily accessible experimental setting for testing "Chern-Simons physics"
and most characteristic phenomena in planar
gauge theories, like topological photon masses, fractional statistics and
Polyakov's confinement mechanism.
Secondly, the actual observation of the oblique confinement
phase of the theory \rqh \ in quantum Hall experiments suggests
that its "sister theory" \mpr \ is indeed a strong candidate for an
effective field theory of quasi-planar superconductivity. Indeed, the
physical mechanism leading to superconductivity in this latter model is
exactly the same mechanism which is responsible for the formation of the
quantum Hall fluids in the former model.

\bigbreak\bigskip\bigskip
\centerline{{\bf Acknowledgements}}
\nobreak \noindent
We would like to thank G. Semenoff for suggesting to us a study of lattice
gauge theories with a mixed Chern-Simons term and
for several enlightning discussions. Moreover, we thank R. Fazio and
G. Sch\"on for discussions on the physical settings of
experiments on Josephson junction arrays.

\listrefs
\end